\documentclass[12pt]{iopart}

\usepackage{graphicx}
\usepackage{textcomp}
\usepackage{siunitx}
\usepackage{xr}
\usepackage{xr-hyper}
\usepackage{url}

\begin{document}

\title[Improving the electrical conductivity of Pt nanowires]{Improving the electrical conductivity of Pt nanowires deposited by focused electron beam induced deposition using thermal annealing}

\author{Rajendra Rai$^1$, Ujjwal Dhakal$^2$,
Binod DC$^1$, and Yoichi Miyahara$^{1,2}$}

\address{$^1$ Materials Science, Engineering and Commercialization program, Texas State University, San Marcos, Texas 78666, USA}
\address{$^2$ Department of Physics, Texas State University, San Marcos, Texas 78666, USA}

\ead{yoichi.miyahara@txstate.edu}

\vspace{10pt}
\begin{indented}
\item[]\today
\end{indented}

\begin{abstract}
We investigated the electrical conductivity of platinum nanowires with heights ranging from 2\,\si{nm} to 200\,\si{nm}, deposited by focused electron beam induced deposition (FEBID).
Post-deposition processing was employed to enhance the electrical conductivity of the platinum nanowires.
Thermal annealing of as-deposited nanowires in air at 225\,$^{\circ}$C for 4 hours increased electrical conductance by up to five orders of magnitude.
After annealing, 22.5\,$\mathrm{\mu m}$-long nanowires with a height of 36\,nm exhibited resistances of approximately 10\,k$\Omega$.
This nanowire underwent a reduction in height to one-quarter of its original value, a reduction in width to one half, and a reduction in cross-sectional area by approximately one order of magnitude.
The platinum-to-carbon weight ratio increased from 35:65 to 85:15.
The electrical resistance decreased monotonically as temperature was lowered from room temperature to 100\,\unit{mK}, confirming that annealed FEBID platinum nanowires are promising building blocks for nanoelectronic devices operating at millikelvin temperatures.
\end{abstract}

% \vspace{2pc}
% \noindent{\it Keywords}: XXXXXX, YYYYYYYY, ZZZZZZZZZ

% \submitto{\JPA}

% \maketitle

% \ioptwocol

\section{Introduction}
Focused electron beam induced deposition (FEBID) enables the direct fabrication of nanostructures in a scanning electron microscope (SEM) without lithographic processing steps \cite{huthFocusedElectronBeam2018c}.
Electrically conducting nanowires (NWs) deposited by FEBID have found applications in nanoelectronic devices, serving as electrical leads and enabling single-electron transistors (SETs) \cite{durraniElectronTransportRoom2017}.
FEBID is well suited for the direct-write fabrication of SETs, where nano-granular platinum islands exhibit tunable tunnel coupling through post-growth irradiation \cite{primaDirectwriteSingleElectron2019}.
However, FEBID-deposited structures typically contain significant impurities originating from organometallic precursor molecules.
For platinum—a material commonly deposited by FEBID—a substantial fraction of carbon is often present, rendering the FEBID-deposited structures electrically non-conducting \cite{fanAnnealingEffectPlatinumincorporated2014}.
Mulders et al.\ reported that the platinum deposits produced via FEBID using MeCpPtMe$_{3}$ (Me: methyl, Cp: cyclopentadienyl) precursor contain only 16\,\% platinum \cite{muldersElectronBeamInduced2010}.
The resulting deposits comprise nanometer-sized grains of face-centered cubic (fcc) platinum embedded in a carbon matrix, with inter-grain spacing on the order of $\sim1$\,nm \cite{deteresaOriginDifferenceResistivity2009}.
Consequently, FEBID-deposited platinum forms granular metals in which electrical transport involves diffusive charge transport within Pt grains and thermally assisted tunneling between grains \cite{huthFocusedElectronBeam2018c}.

\begin{table}[hhhh]
\centering
\caption{Comparison of electrical conduction of Pt nanowires deposited by FEBID and FIBID}
\label{table : comparison}
\footnotesize
\begin{tabular}{>{\centering\arraybackslash}m{1.9cm}>{\centering\arraybackslash}m{2cm}>{\centering\arraybackslash}m{2cm}>
{\centering\arraybackslash}m{1.2cm}>{\centering\arraybackslash}m{1.2cm}>
{\centering\arraybackslash}m{2cm}>{\centering\arraybackslash}m{3cm}}
\hline
Source & Post-treatment & Beam voltage and current & Height (nm) & Width (nm) & Room-temperature resistivity ($\mu\Omega$m) & Temperature dependence \\
\hline
& & & & & & \\
This work (FEBID) & Annealing in air at 225$^\circ$C & 2\,kV, 43\,pA & 31 & 200 & 3.0 & Metallic down to 100\,mK \\
& & & & & & \\
Gopal et al.\ (FEBID) (2004)~\cite{gopalRapidPrototypingSiteSpecific2004} & Rapid annealing in nitrogen at 600$^\circ$C  & 10\,kV, 2000\,pA & N/A & 250 & $\approx 10$ & N/A \\
& & & & & & \\
Botman et al.\ (FEBID) (2006)~\cite{botmanPurificationPlatinumGold2006} & Annealing in oxygen-containing gas at 300$^\circ$C (10 min) & 20\,kV, 620\,pA & 150 & 300 & $1.4 \times 10^2$ & N/A \\
& & & & & & \\
De Teresa et al., 2009 (FEBID) \cite{deteresaOriginDifferenceResistivity2009} & In-situ electron irradiation & 1$\sim$30\,kV, 400$\sim$9500\,pA & 160 & 1000 & $10^{5}$ & Semiconducting \\
& & & & & & \\
Porrati et al., 2011 (FEBID) \cite{porratiTuningElectricalConductivity2011a} & Electron irradiation & 5\,kV, 1600\,pA & 80 & 1000 & 7.8 & Metallic down to 1.8\,K \\
& & & & & & \\
\hline
\end{tabular}
\end{table}

Table \ref{table : comparison} summarizes previous experiments aimed at improving the conductivity of FEBID-Pt deposits through in-situ or post-processing methods such as electron beam irradiation and thermal annealing \cite{deteresaOriginDifferenceResistivity2009,
gopalRapidPrototypingSiteSpecific2004,   
botmanPurificationPlatinumGold2006,
porratiTuningElectricalConductivity2011a}.
Gopal et al. reported that rapid annealing in nitrogen at 600\,$^\circ$C reduces 
the room-temperature resistivity by a factor of 20 \cite{gopalRapidPrototypingSiteSpecific2004}.
Botman et al.\ demonstrated that annealing in an oxygen atmosphere increases the Pt content \cite{botmanPurificationPlatinumGold2006}.
Porrati et al.\ reported a three-order-of-magnitude increase in conductivity after 30 minutes of electron irradiation, achieving a resistivity of 7.8\,$\mu \Omega$m.
The enhancement was attributed to graphitization (the formation of graphite crystals), carbon removal, and percolation (the formation of conductive pathways in composite materials) \cite{porratiTuningElectricalConductivity2011a}.
De Teresa et al.\ contrasted different conduction mechanisms in Pt deposits produced by focused ion beam induced deposition (FIBID) and FEBID.
They reported that while FIBID-deposited Pt exhibits metallic conduction, FEBID-deposited Pt shows variable-range hopping conduction below 200\,K \cite{deteresaOriginDifferenceResistivity2009}.
Fang et al.\ annealed Pt-incorporated nanowires in a pure oxygen atmosphere with a flow rate of 30\,sccm.
They found that although annealing increased the Pt content, it caused collapse of the deposited structure.
Despite the observed increase in Pt content after annealing, the nanowires exhibited structural cracking and, consequently, no measurable current was detected \cite{fanAnnealingEffectPlatinumincorporated2014}.
Huth et al.\ linked electron irradiation to enhanced inter-grain coupling \cite{huthFocusedElectronBeam2012}.
This improvement was attributed to a matrix transformation from amorphous carbon to nanocrystalline graphite, which strengthens electronic coupling between Pt grains.
They observed that temperature-dependent transport evolves from correlated variable-range hopping in as-deposited nanostructures to a universal logarithmic dependence of conductivity after intense electron irradiation, indicating the emergence of a granular Fermi-liquid regime at low temperatures \cite{Sachser2011PRL}.

In this work, we deposited Pt nanowires of varying dimensions by FEBID and investigated the effects of thermal annealing on their dimensions and electrical conductivity.
We found that thermal annealing at $225^{\circ}$C in air reduced the carbon content and decreased the electrical resistance by up to five orders of magnitude.
Atomic force microscopy reveals that the height decreases to 25\,\% or less of that of the as-deposited nanowires.
The cross-sectional area is reduced by nearly an order of magnitude, while the width decreases to about half of its value before annealing.
The temperature dependence of the resistance of the annealed Pt nanowires confirms metallic conduction down to 100\,mK.
The annealed Pt nanowires achieve resistivity as low as $3.0\,\mu\Omega$m, which is 28 times higher than the bulk value.
This is an important step towards building reliable nanoelectronic devices that operate at millikelvin temperatures, such as single-electron transistors fabricated by FEBID.
Such highly conducting nanowires will play a vital role in further miniaturization of electronic devices.

\section{Experimental Methods}

\subsection{Electrode fabrication}
\begin{figure}[htbp]
\centering
\includegraphics[width=0.99\textwidth]{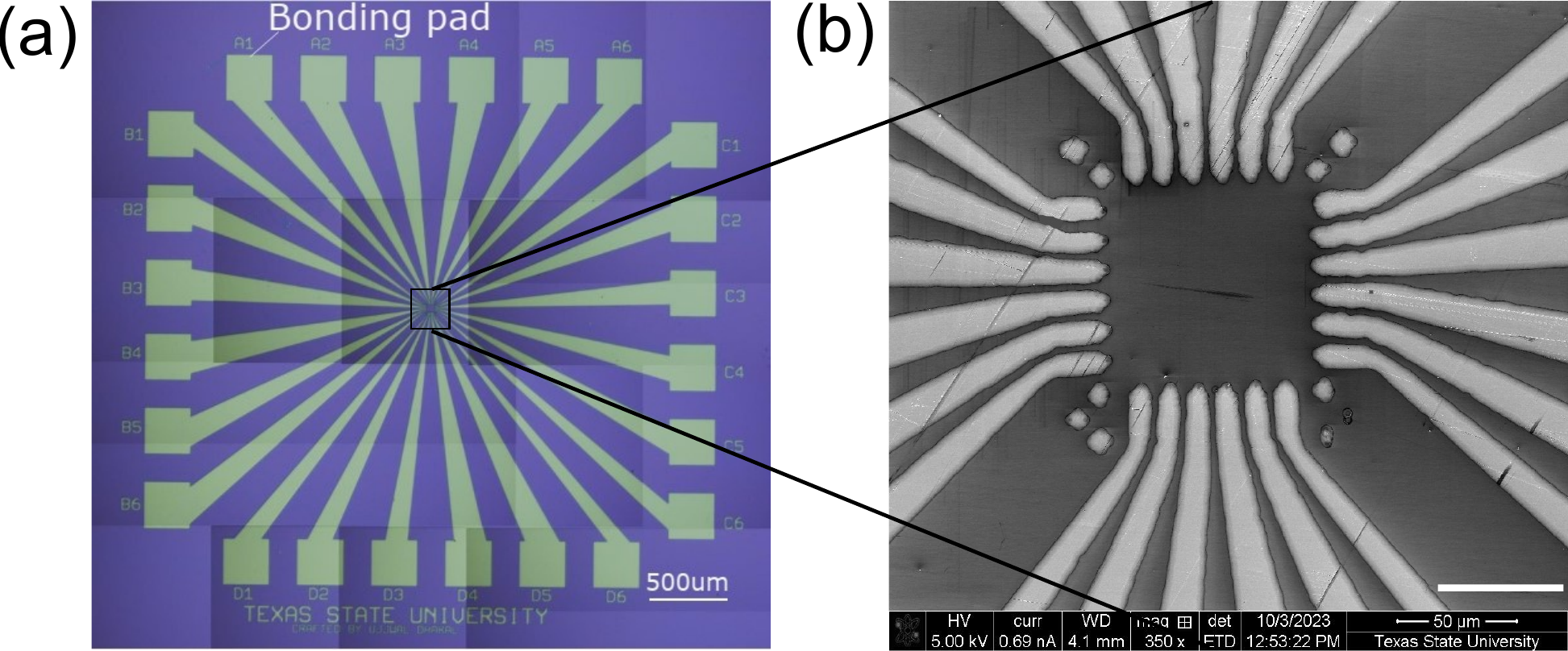}
\caption{(a) Optical and (b) SEM micrographs of gold electrodes patterned on a SiO$_2$/Si substrate. There are twenty-four electrodes.
Scale bar in (b) is 50\,$\mu$m.}
\label{fig:microelectrodes}
\end{figure}

We deposited gold electrodes on SiO$_2$/Si substrates for two-probe resistivity measurements of the nanowires.
Gold was chosen due to its electrical conductivity and chemical stability.
In addition, it ensures low contact resistance with the nanowires.
The microelectrode array of gold was fabricated on a 285\,nm-thick SiO$_2$ layer grown on a heavily doped p-type silicon wafer using standard photolithography and lift-off techniques.
As shown in Fig.~\ref{fig:microelectrodes}, the substrate contains 24 patterned gold electrodes (40\,nm-thick gold with a 10\,nm-thick chromium adhesion layer).
The minimum gap between the electrodes is 2\,$\mu$m.
Each electrode is designed to connect to FEBID-deposited Pt nanowires for subsequent experiments.

\subsection{Deposition of platinum using the FEBID technique}
\label{startsample}
We performed FEBID deposition of Pt nanowires in a scanning electron microscope (Helios NanoLab 400 Dual Beam, FEI), using $(\mathrm{CH}_3)_{3}$Cp$\mathrm{Pt}(\mathrm{CH}_{3})$ ((trimethyl)(methyl)(cyclopentadienyl)platinum(IV)) as a precursor.
The nominal height of each deposited Pt nanowire was specified by the ``Z size''  parameter in the patterning software to control the nanowire height.
Note that the nominal height (``Z size'') may differ from the actual measured height.
We deposited Pt nanowires with nominal heights ranging from 2\,nm to 200\,nm and lengths from 1\,$\mu$m to 60\,$\mu$m.

We used two different deposition conditions depending on the nanowire length.
For nanowires 2\,$\mu$m in length or shorter, we used an acceleration voltage of 2\,kV, a beam current of 43\,pA, and a pitch of 1\,nm.
For longer nanowires (up to 60\,$\mu$m), we employed a higher acceleration voltage of 5\,kV, a beam current of 690\,pA, and a pitch of 22.5\,nm.
We used the higher current and voltage to reduce the deposition time for longer nanowires.
At \(5~\mathrm{kV}\) and \(690~\mathrm{pA}\), a nanowire with a length of approximately \(15.08~\mu\mathrm{m}\) and a nominal height of \(133~\mathrm{nm}\) was deposited within \(21~\mathrm{s}\).
In contrast, under lower beam conditions of \(2~\mathrm{kV}\) and \(43~\mathrm{pA}\), a nanowire of similar length (\(15.48~\mu\mathrm{m}\)) and a significantly smaller nominal height of \(2.64~\mathrm{nm}\) required \(20~\mathrm{s}\) for deposition.
Based on this comparison, the deposition at \(5~\mathrm{kV},~690~\mathrm{pA}\) is approximately 50 times faster in terms of vertical growth rate than that achieved at \(2~\mathrm{kV},~43~\mathrm{pA}\).

For all depositions, the dwell time and the relative interaction diameter were 1.4\,ms and 150\,\%, respectively.
The number of passes that the electron beam drew over a nanowire was changed according to deposition parameters such as height, beam current, and acceleration voltage.
For post-annealing, we annealed the samples at $225^{\circ} \mathrm{C}$ for 4 hours in air using a laboratory oven (HUMBOLDT, 20 GC Lab Oven, see Supporting Figure S1).

\subsection{Characterization of Pt nanowires}
We measured the $I$-$V$ characteristics of the Pt nanowires using a Keysight B1500A semiconductor analyzer connected to a probing station at room temperature in air.
We used a two-probe method, as the contact resistance, including the resistance of the microelectrodes, was found to be 80\,$\Omega$, which is negligible compared to the nanowire resistance ($> 2$\,k$\Omega$).
During measurements, a voltage was swept between two probes and the resulting current through the nanowire was recorded.
This process was repeated for various voltage ranges; typically $I$-$V$ curves were measured from $-100\,\si{mV}$ to 100\,\si{mV}.

We measured the temperature dependence of the two-point resistance across Pt nanowires using a Physical Property Measurement System (PPMS, Dynacool, Quantum Design).
Samples were wire-bonded from the bonding pads on the microelectrodes to the PPMS puck using gold wires to establish electrical contacts.
Resistance was measured at a constant current of $9.25 \times 10^{-7}$\,A by recording the voltage drop across the Pt nanowires.
An average over 25 measurements was recorded at each temperature.

For atomic force microscopy (AFM) topography imaging of the nanowires, we used a commercial atomic force microscope (Smart AFM, Horiba) operated in tapping mode.
Platinum-coated AFM cantilevers (OPUS240AC-PP, Mikromasch) with a nominal resonance frequency of 70\,kHz and a nominal force constant of 2\,N/m were used.
The free oscillation amplitude was set to 20\,nm.

Energy-dispersive X-ray spectroscopy (EDS) analysis of these nanowires was performed at 10\,keV with the beam current set to the highest possible value to obtain high signal intensities in the spectra, typically greater than 1.4\,nA.

\section{Results and Discussion}

\subsection{Effect of annealing on nanowire composition}
\begin{figure}
\centering
\includegraphics[width=\textwidth]{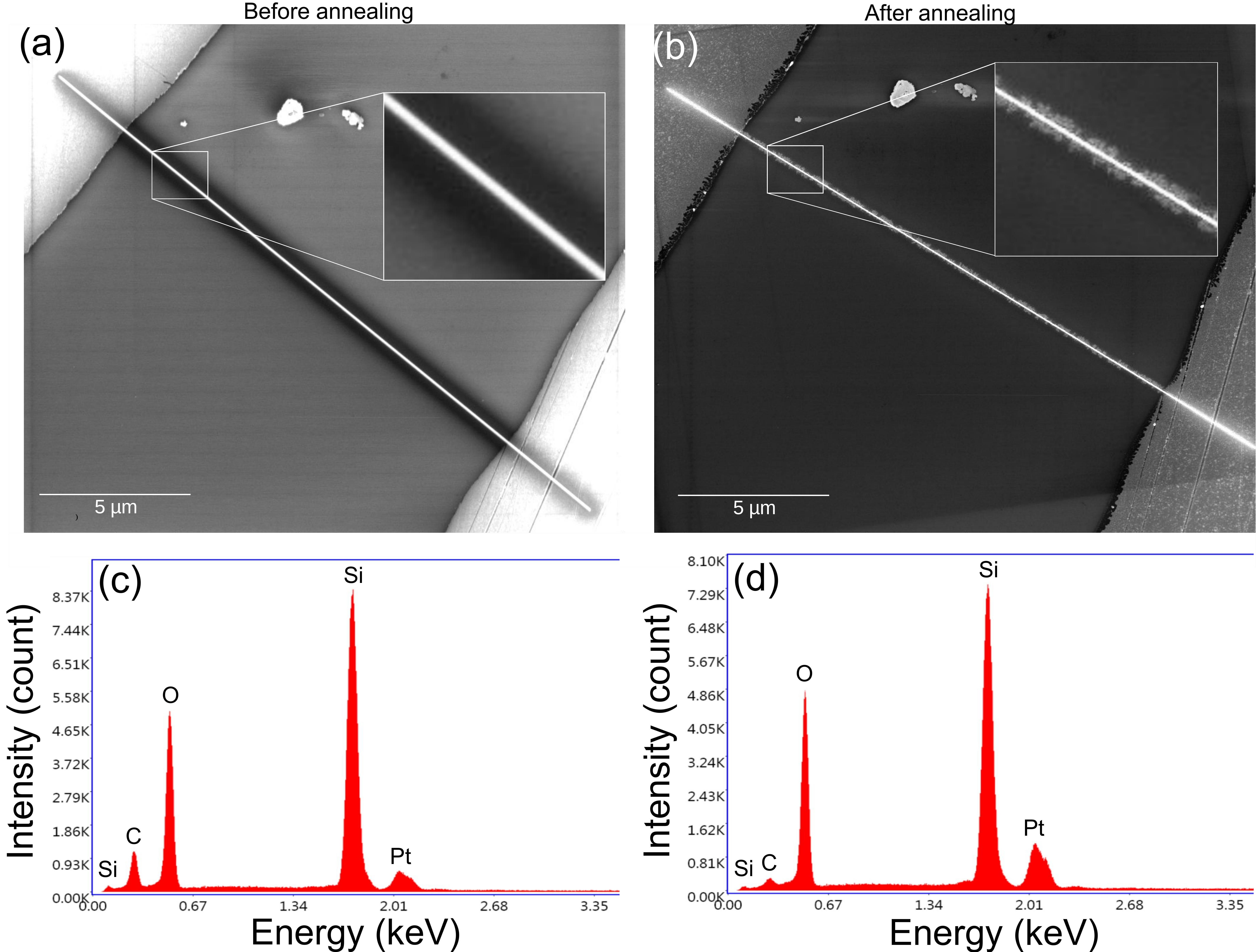}
\caption{SEM images of a FEBID-deposited nanowire with a nominal height of 133\,nm:
(a) before annealing (the nanowires are freshly deposited and the image is taken immediately afterward) and
(b) after annealing (4 hours at 225$^{\circ}$C in air).
The insets in (a) and (b) are magnified views.
EDS spectra taken for a different nanowire with a length of 5\,$\mu$m and a height of 50\,nm (c) before and (d) after annealing.
The nanowire was deposited at 5\,kV and 690\,pA.}
\label{fig:SEM_BA_AA}
\end{figure}

Figure \ref{fig:SEM_BA_AA} shows SEM images of a Pt nanowire with a nominal height of 133\,nm before (a) and after (b) annealing.
The nanowire was deposited at an operating voltage of 5\,kV and a beam current of 690\,pA.
The number of passes was one and the dwell time was 1.4\,ms.
We observe a clear difference between the SEM images taken before and after annealing.
After annealing, the width of the nanowire appears smaller, and fuzzy granular structures are visible along the sides of the nanowire (see the inset in Fig.~\ref{fig:SEM_BA_AA}(b)).

To confirm compositional changes in the nanowire induced by annealing, we performed EDS measurements on a different nanowire with a length of 5\,$\mu$m and a nominal height of 50\,nm deposited at 2\,kV and 43\,pA.
The EDS spectra of the nanowire before and after annealing are shown in Fig.~\ref{fig:SEM_BA_AA}(c) and (d), respectively.

\begin{table}[h!]
\caption{Results of EDS analysis: weight and atomic percentages of a Pt nanowire deposited at 2\,kV and 43\,pA with a nominal height of 50\,nm before and after annealing.}
\label{tab:percentages}
\footnotesize
\centering
\begin{tabular}{lrrrr}
\hline
Element & \multicolumn{2}{c}{Before annealing} & \multicolumn{2}{c}{After annealing} \\
& Weight \% & Atomic \% & Weight \% & Atomic \% \\
\hline
Carbon (C) & 6.30 & 11.49 & 4.22 & 9.37 \\
Platinum (Pt) & 1.61 & 0.18 & 24.22 & 3.31 \\
Silicon (Si) & 64.21 & 50.12 & 44.71 & 42.50 \\
Oxygen (O) & 27.89 & 38.21 & 26.85 & 44.81 \\
\hline
\end{tabular}
\end{table}

Table~\ref{tab:percentages} summarizes the EDS compositional analysis results of the nanowire before and after annealing.
The relative ratio of carbon to platinum changes from 80\,\% to 20\,\% by weight before annealing to 5\,\% carbon and 85\,\% platinum after annealing.
The presence of Si and O is attributed to the SiO$_{2}$ substrate on which these nanowires were deposited.

\subsection{Effect of annealing on nanowire dimensions}
\begin{figure}
\centering
\includegraphics[ width=\textwidth]{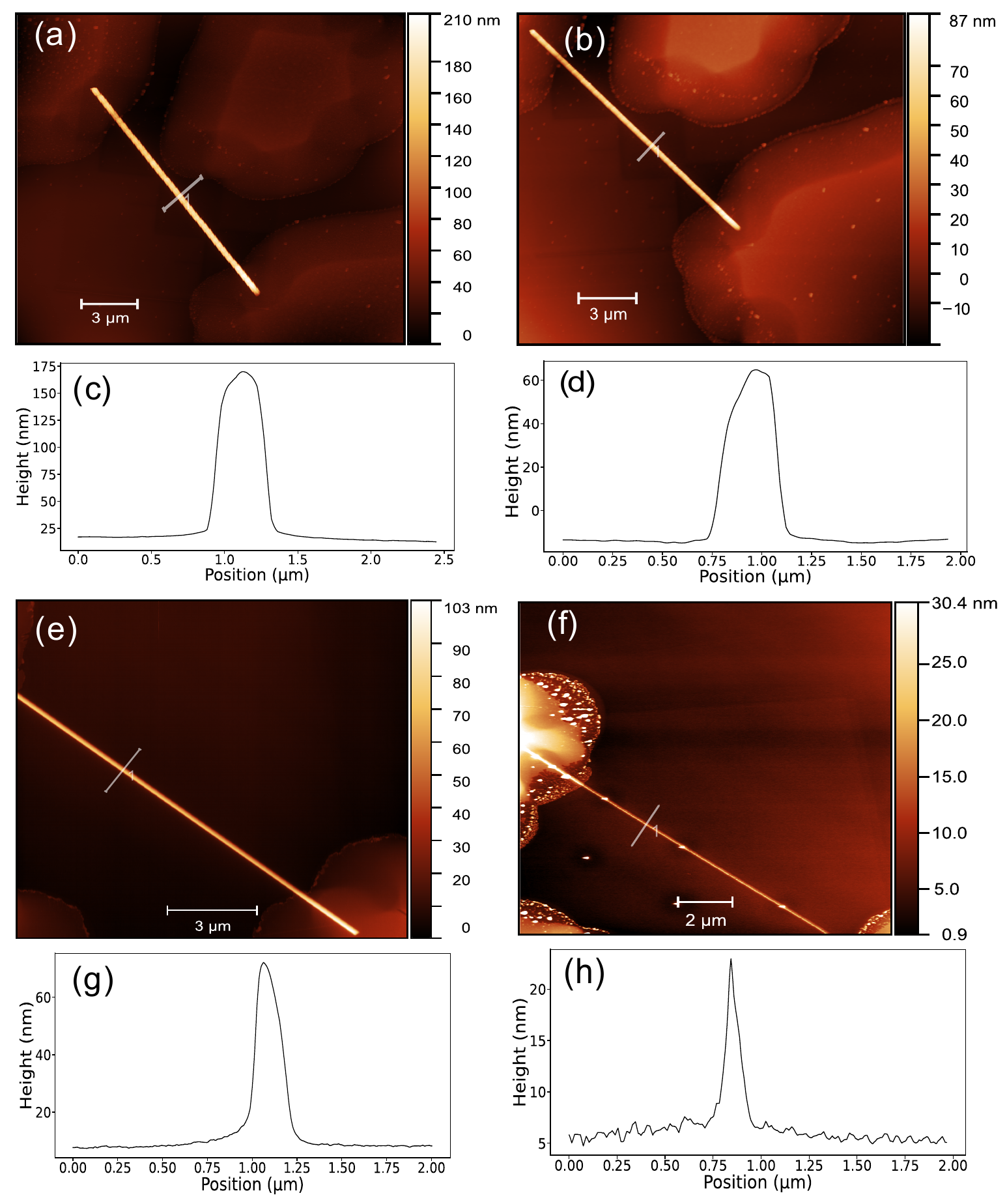}
\caption[AFM images of a 133\,nm wire]{AFM topography images of Pt nanowires deposited under different beam conditions:
(a) at 5\,kV acceleration voltage and 690\,pA beam current before annealing and
(b) after annealing.
(c) and (d) show the corresponding line profiles of the nanowires indicated by the white lines in (a) and (b).
Similarly, AFM topography images of a Pt nanowire with a nominal height of 20\,nm deposited at 2\,kV acceleration voltage and 43\,pA beam current are shown
(e) before annealing and (f) after annealing, with their respective line profiles presented in (g) and (h).}
\label{fig: afm_130nm_ba_aa}
\end{figure}

\begin{figure}[h]
\centering
\includegraphics[width=0.97\textwidth]{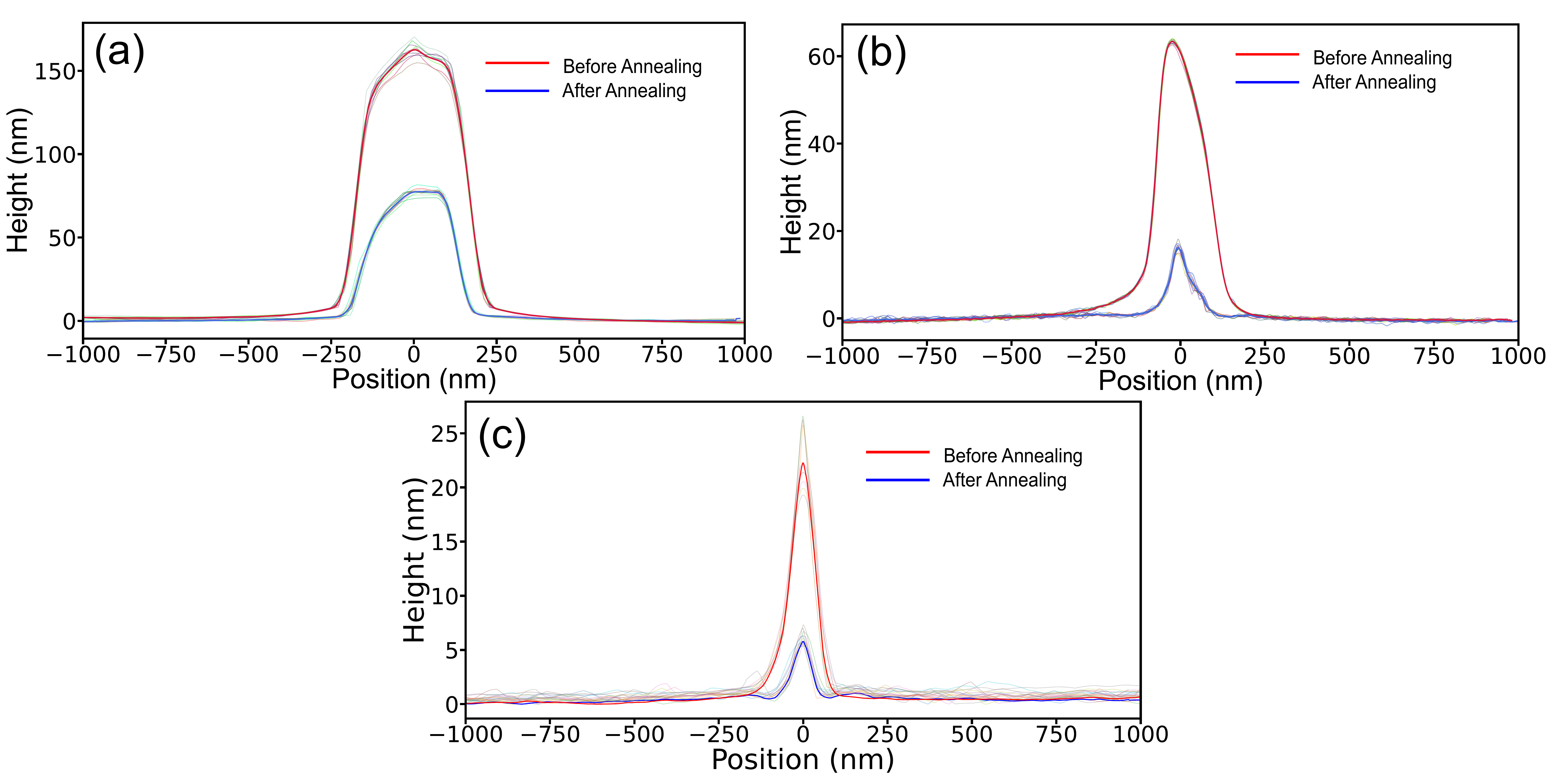}
\caption{
Multiple height profiles extracted from the AFM topography images of Pt nanowires.
The thick red (blue) line represents the averaged topography of all profiles before (after) annealing.
These plots reveal how the nanowire topography evolves with annealing.
(a) Pt nanowire deposited at an acceleration voltage of 5\,kV and 690\,pA beam current with a nominal height of 133\,nm and dwell time of 1.4\,ms (Fig.~\ref{fig: afm_130nm_ba_aa}a and b).
(b) Pt nanowire deposited at an acceleration voltage of 2\,kV and 43\,pA beam current with nominal height 20\,nm and dwell time of 1.4\,ms (Fig.~\ref{fig: afm_130nm_ba_aa}e and f).
(c) Pt nanowire deposited at an acceleration voltage of 2\,kV and 43\,pA beam current with a dwell time of 1.4\,ms and a single pass (see Supporting Fig.~S4).}
\label{fig:dwell_time_5kV}
\end{figure}

Figure~\ref{fig: afm_130nm_ba_aa} presents AFM topography images of two Pt nanowires before and after annealing.
The two nanowires were deposited under different conditions.
The nanowire shown in panel (a) (before annealing) and panel (b) (after annealing) was deposited at 5\,kV and 690\,pA with a nominal height of 133\,nm, using a pitch of 1\,nm, a dwell time of 1.4\,ms, and a single pass.
The measured height of the as-deposited nanowire was 161\,nm (Fig.~\ref{fig: afm_130nm_ba_aa}(c)), which decreased to 77\,nm (Fig.~\ref{fig: afm_130nm_ba_aa}(d)) after annealing at 225\,\textdegree C for 4 hours in air.
To further quantify these changes, multiple topography line profiles were extracted from the AFM images (Fig.~\ref{fig: afm_130nm_ba_aa}(a) and (b)), as shown in Fig.~\ref{fig:dwell_time_5kV}(a).
The extracted average width and cross-sectional area are 300\,nm and $5.3\times10^4$\,nm$^2$, respectively.
After annealing, these values decreased to 77\,nm in height, 250\,nm in width, and $2.1\times 10^4$\,nm$^2$ in cross-sectional area.
For details of the extraction of width and cross-sectional area, see Section S5 in the supporting information.

Similarly, Fig.~\ref{fig: afm_130nm_ba_aa}(e) (before annealing) and (f) (after annealing) show AFM topography images of a Pt nanowire deposited at 2\,kV and 43\,pA with a nominal height of 20\,nm, a pitch of 1\,nm, a dwell time of 1.4\,ms, and three passes.
Figure~\ref{fig: afm_130nm_ba_aa}(g) and (h) show the corresponding height profiles.
The measured height before annealing was approximately 60\,nm (Fig.~\ref{fig: afm_130nm_ba_aa}(g)), which decreased to 13\,nm after annealing at 225\,\textdegree C for 4 hours in air (Fig.~\ref{fig: afm_130nm_ba_aa}(h)).

To quantify the results more precisely, multiple topography profiles were extracted from the AFM topography of the nanowire (Fig.~\ref{fig:dwell_time_5kV}(b)).
The as-deposited nanowire exhibited an average height of 62\,nm, 
a width of 160\,nm, and a cross-sectional area of $1.0 \times 10^4\,\mathrm{nm}^2$.
After annealing, the height, width, and cross-sectional area of the nanowire decreased to 16\,nm, 95\,nm, and $1.3 \times 10^3\,\mathrm{nm}^2$, respectively.

\begin{figure}[h]
\centering
\includegraphics[width=\textwidth]{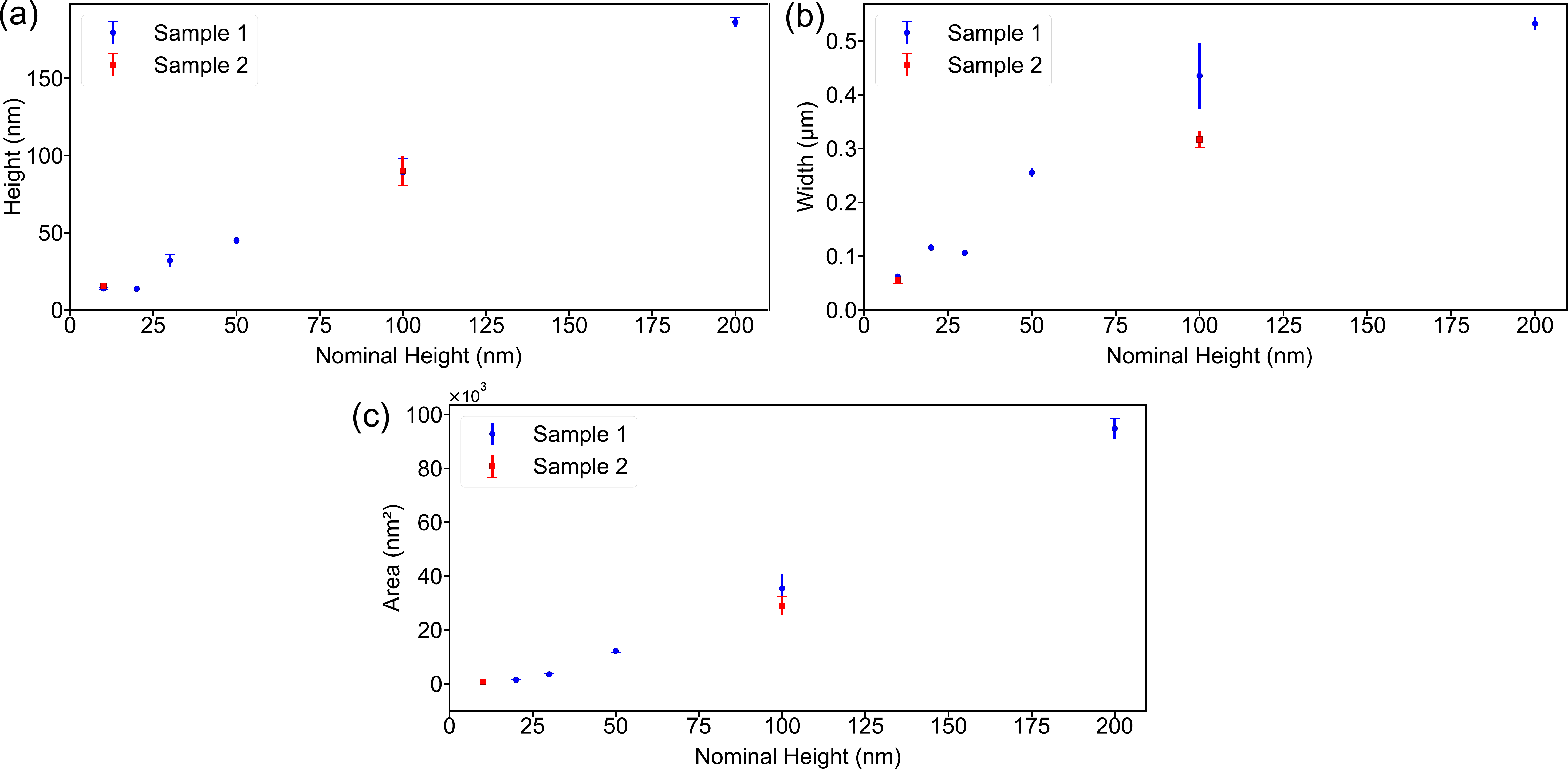}
\caption{
Variation of nanowire dimensions with nominal height after annealing.
(a) Measured height vs nominal height  after annealing.
(b) Width vs nominal height vs  after annealing.
(c) Cross-sectional area vs nominal height after annealing.
Sample 1 and Sample 2 refer to two different sets of Pt nanowires with the same nominal dimensions but deposited on different substrates (for calculation details see Section S5).}
\label{fig:Nominal_height_structures}
\end{figure}

Table~\ref{tab:NW_dimensions} summarizes the deposition and annealing parameters of the Pt nanowires studied.
\begin{table}[ht]
\centering
\caption{Deposition and annealing parameters of Pt nanowires with their measured dimensions before and after annealing.}
\label{tab:NW_dimensions}
\footnotesize
\setlength{\tabcolsep}{4pt}
\renewcommand{\arraystretch}{1.15}
\resizebox{\textwidth}{!}{%
\begin{tabular}{cccccccccc}
\hline
Voltage & Beam & Nominal &  & \multicolumn{3}{c}{Before annealing} & \multicolumn{3}{c}{After annealing} \\
(kV) & current (pA) & height (nm) & Passes & Height (nm) & Width (nm) & Area (nm$^2$) & Height (nm) & Width (nm) & Area (nm$^2$) \\
\hline
5 & 690 & 133 & 1 & 161 & 300 & $5.3\times 10^{4}$ & 77 & 250 & $2.1\times 10^{4}$ \\
2 & 43 & 20 & 3 & 62 & 160 & $1.0\times 10^{4}$ & 16 & 95 & $1.3\times 10^{3}$ \\
2 & 43 & 6 & 1 & 21.5 & 91 & $1.9 \times 10^3$ & 5.3 & 68 & $3.4 \times 10^{2}$ \\
\hline
\end{tabular}}
\end{table}
These results provide a comparison of the dimensional changes under different deposition conditions.
At 5\,kV and 690\,pA with a nominal height of 133\,nm, the nanowire height decreased by about 52\,\%, the width by 17\,\%, and the cross-sectional area by 60\,\%.
Under 2\,kV, 43\,pA with a nominal height of 20\,nm (three passes), 
the reductions were more pronounced: height decreased by 74\,\%, width by 41\,\%, and area by 87\,\%.
For the 2\,kV, 43\,pA single-pass condition with a nominal height of 6\,nm, the nanowire exhibited the most significant thinning, 
with height reduced by 75\,\%, width by 25\,\%, and area by 82\,\%.
These results indicate that dimensional reductions were greater 
for nanowires deposited at 2\,kV than for those deposited at 5\,kV.
% In most cases, the measured height of the annealed nanowires (see Fig.~S2 for AFM images) became close to the nominal height.
Moreover, for nanowires deposited at 2\,kV and 43\,pA, 
the measured height before annealing was larger than the nominal value, 
but after annealing, it closely matched the nominal height.
Supplementary Figure S3 shows the height profiles of the annealed nanowires
with various nominal heights, confirming this trend.
Figure~\ref{fig:Nominal_height_structures} shows the relations between 
the dimension of the nanowires (height, width and cross-sectional area) 
and nominal height taken from Fig.~S3.

Figure~\ref{fig:dwell_time_5kV}(c) presents the height profiles obtained across a Pt nanowire deposited at 2\,kV and 43\,pA with a dwell time of 1.4\,ms and a single pass (the topography image is shown in Fig.~S4).
In single-pass deposition, the inherent shrinkage of nanowire dimensions following annealing can be directly observed
because there is no additional width broadening
that can happen in multiple pass writing.
The average dimensions of the as-deposited nanowire are a height of 21.5\,nm, a width (FWHM) of 91\,nm, and a cross-sectional area of 1897\,nm$^{2}$.
The dimensions of the annealed nanowire decreased to a height of 5.3\,nm, a width of 68\,nm, and a cross-sectional area of 339\,nm$^{2}$, clearly showing significant shrinkage in all dimensions.
The profiles demonstrate that the nanowire undergoes substantial morphological changes as a result of annealing.
Specifically, the cross-sectional area of the structure is reduced by up to a factor of 10, the height decreases by a factor of 5, and the width shrinks by nearly half.

\subsection{Effect of annealing on $I$-$V$ curves}
\begin{figure}[h]
\centering
\includegraphics[width=0.99\textwidth]{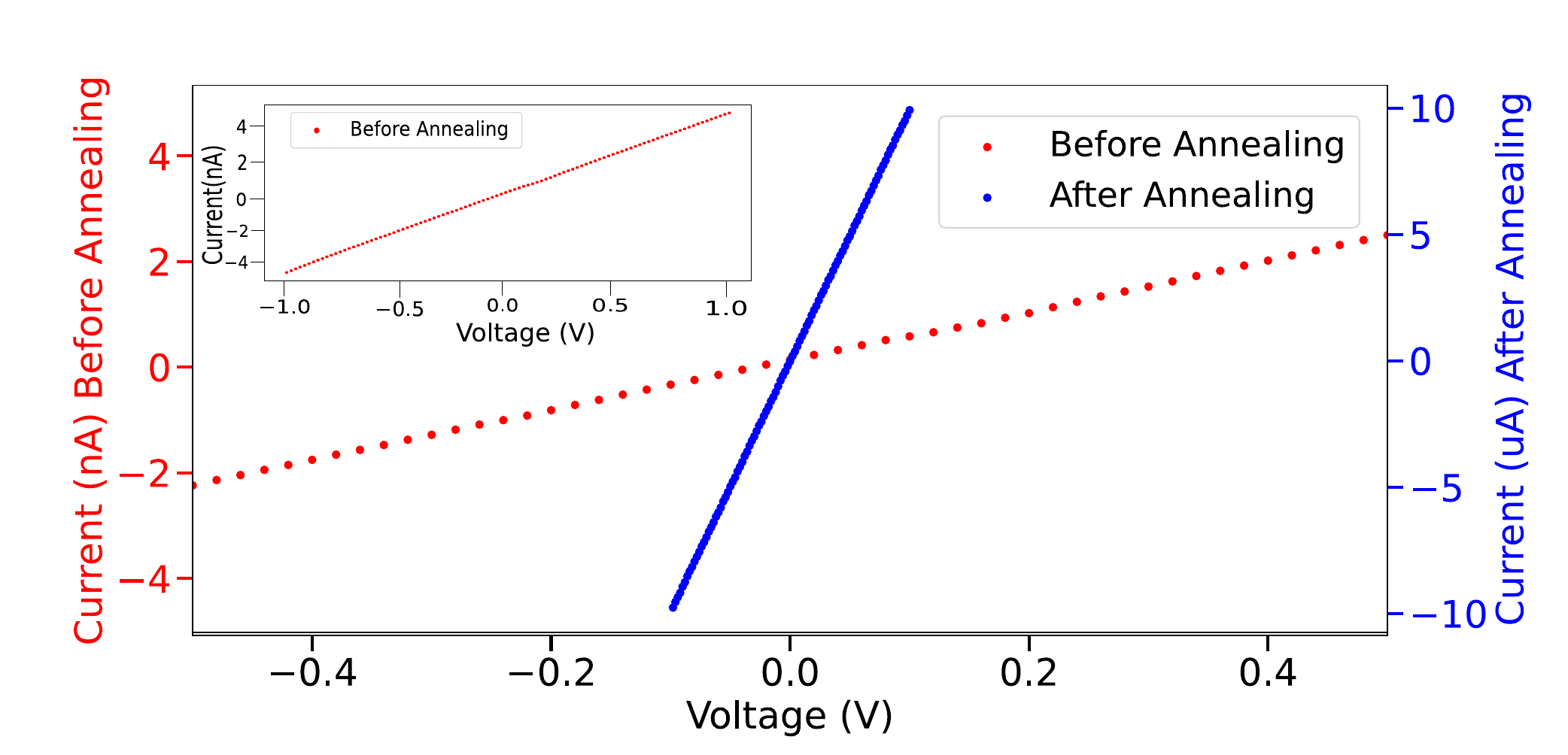}
\caption{Effect of annealing on $I$-$V$ curves of a 36.5\,nm-thick nanowire before (red) and after (blue) annealing.
The annealing was done for 4 hours at 225$^{\circ}$C.
This is the $I$-$V$ characteristic of the same nanowire (nominal height 133\,nm) shown in Fig.~\ref{fig:SEM_BA_AA}.
The inset shows the conducting behavior of the nanowire before annealing.}
\label{fig:Resistance_BA_AA}
\end{figure}

Figure~\ref{fig:Resistance_BA_AA} shows $I$-$V$ curves of a Pt nanowire deposited at 5\,kV and 690\,pA with a nominal height of 133\,nm before (blue) and after (red) annealing.
The height of the nanowire was reduced to 40\,nm after annealing.
The resistance of this nanowire was 250\,M$\Omega$ before annealing.
The $I$-$V$ curves show linear relationships both before and after annealing.
After annealing at 225\,$^\circ$C for 4 hours, the resistance decreased to 10\,k$\Omega$.
This substantial reduction in resistance after annealing is a common trend observed in most of the nanowires.

\subsection{Resistance vs dimension of nanowires}
\begin{figure}[h]
\centering
\includegraphics[width=0.99\textwidth]{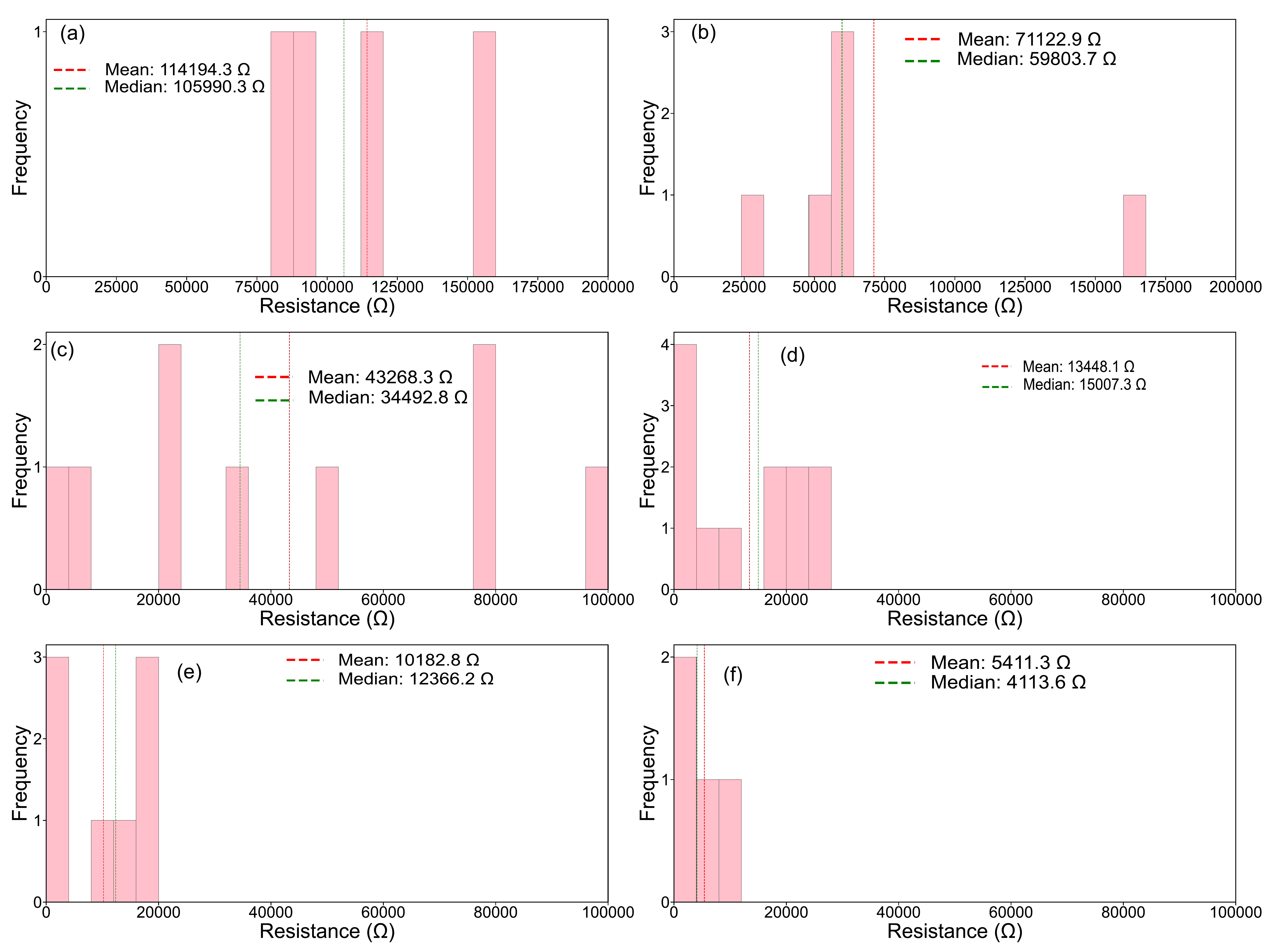}
\caption{Histograms of the resistances of annealed Pt nanowires with different nominal heights:
(a) 10\,nm, (b) 20\,nm, (c) 30\,nm, (d) 50\,nm, (e) 100\,nm, and (f) 200\,nm.
These Pt nanowires were deposited at 2\,kV and 43\,pA under a magnification of $35,000\times$ and annealed for 4.5\,hours.
The pitch value for all these depositions was 1\,nm and the length of the nanowires was 2.1\,$\mu$m.}
\label{fig:Histogram}
\end{figure}

As shown in Fig.~\ref{fig:Resistance_BA_AA}, the resistance of nanowires typically decreases by five orders of magnitude after annealing.
Whereas previous studies reported that annealing led to nanowire fragmentation into discontinuous structures \cite{fanAnnealingEffectPlatinumincorporated2014}, our results demonstrate that the nanowires remained continuous after annealing, resulting in a dramatic improvement in electrical conductivity.

Figure~\ref{fig:Histogram} shows histograms of the resistance measured for annealed nanowires deposited at 2\,kV and 43\,pA with nominal heights of (a) 10\,nm, (b) 20\,nm, (c) 30\,nm, (d) 50\,nm, (e) 100\,nm, and (f) 200\,nm.
The length of all the nanowires was 2.1\,$\mu$m.
The pitch was 1\,nm.
For each histogram, the mean and median of the resistance values are shown.
\begin{figure}[t]
\centering
\includegraphics[width=0.8\linewidth]{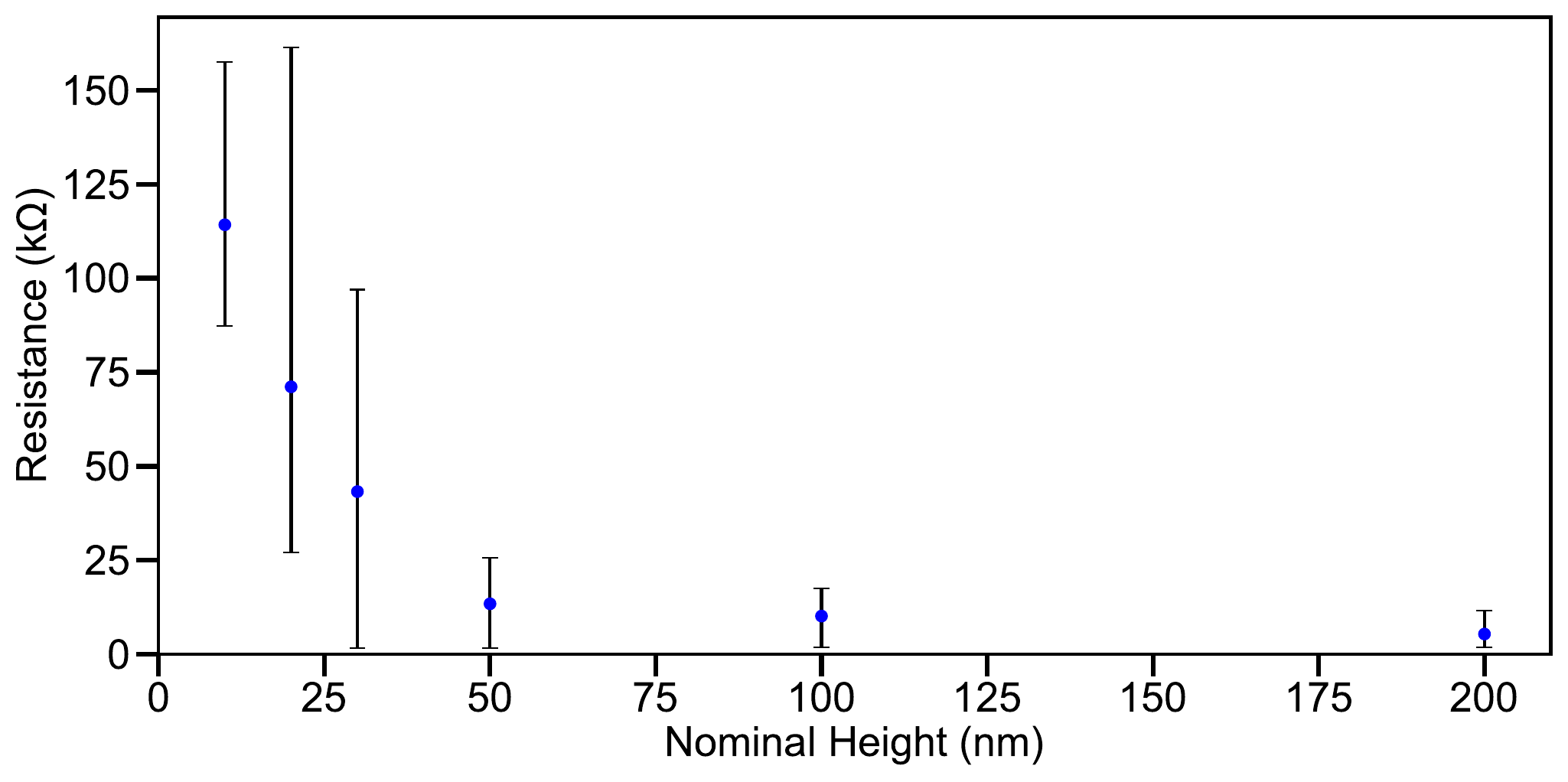}
\caption{Resistance vs nominal height for annealed FEBID-deposited Pt nanowires with nominal heights ranging from 10 to 200\,nm.
Each marker represents the mean resistance at a given nominal height.
For each height, the mean resistance value was calculated, and the lower and upper ends of the error bars correspond to the minimum and maximum resistance values, respectively.}
\label{fig:R_vs_thickness}
\end{figure}
The yield of uniform and morphologically well-defined nanowires is highly sensitive to the focusing quality of the electron beam.
In practice, for deposition at \(2\,\mathrm{kV}\) and \(43\,\mathrm{pA}\) with a working magnification of $35,000\times$, the beam is first precisely focused at $65,000\times$ magnification and then reduced to $35,000\times$ for deposition.
This procedure ensures that the beam remains well focused, which results in nanowires with consistent geometry and smooth morphology.

From the histograms, we plot the resistance vs nominal height in Fig.~\ref{fig:R_vs_thickness}.
The figure reveals that the mean resistance decreases as the nominal height of the nanowires increases.
At the same time, the spread in resistance values decreases for thicker nanowires.
The lowest measured resistance appears to level off once the nominal height exceeds approximately 50\,nm.

\begin{figure}
\centering
\includegraphics[width=\textwidth]{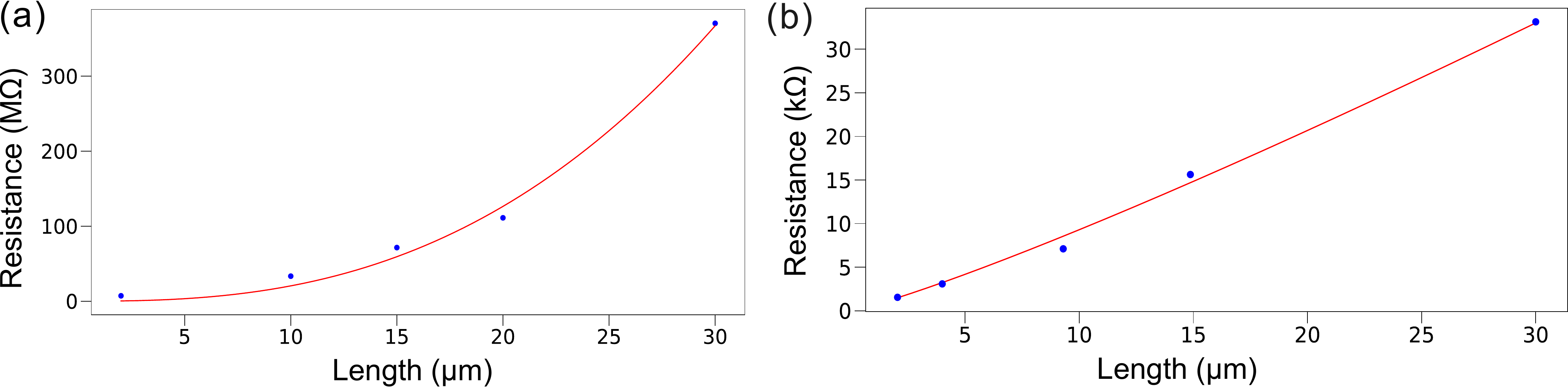}
\caption{
Resistance vs length of the nanowires deposited at 5\,kV, 690\,pA, a dwell time of 1.4\,ms, and a pitch value of 22.50\,nm
(a) before and (b) after annealing.}
\label{fig:ResistanceVSlength}
\end{figure}

Figure~\ref{fig:ResistanceVSlength} shows the relationship between the resistance and the length of a Pt nanowire deposited at 5\,kV and 690\,pA with a nominal height of 133\,nm and a dwell time of 1.4\,ms, (a) before and (b) after annealing.
The experimental data were fitted using a power-law model of the form $R = aL^{b}$, where $R$ denotes the resistance and $L$ the length of the nanowire.
For the as-deposited nanowires, the fitting yielded $a = 4.77 \times 10^{4}$ and $b = 2.6310$, whereas after annealing the parameters were reduced to $a = 652.021$ and $b = 1.1539$.
After annealing, the exponent approaches unity ($b \approx 1$), which indicates that the resistance varies linearly with length, consistent with bulk-like transport behavior.

\subsection{Estimation of volume resistivity}
The volume resistivity of the deposited nanowires was obtained by combining the measured resistance (Fig.~\ref{fig:R_vs_thickness}) with the measured cross-sectional area extracted from AFM topography (Fig.~\ref{fig:Nominal_height_structures}(c)).
Using these AFM-derived values for the area \(A\), together with the measured resistance \(R\) of the nanowires and their known length of \(L = 2.1~\mu\text{m}\), the resistivity, $\rho$, was calculated according to \(\rho = R \times A / L\).

\begin{figure}
\centering
\includegraphics[width=\textwidth]{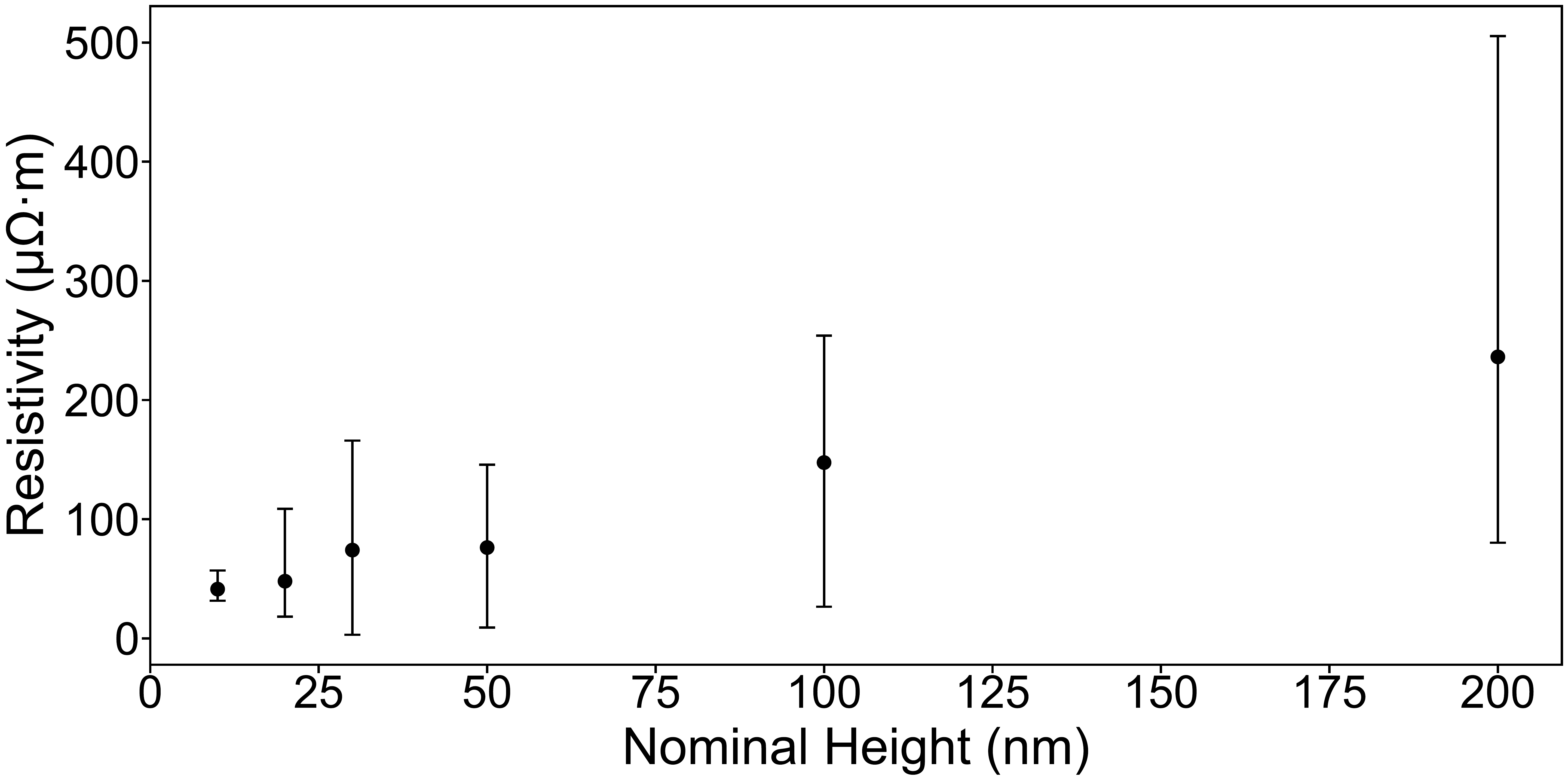}
\caption{Resistivity of annealed Pt nanowires deposited at 2\,kV and 43\,pA vs nominal height of the nanowires.
For each height, a circular marker shows the median value of the resistivity of multiple nanowires.
The error bars represent the minimum and maximum values.
The cross-sectional areas of the nanowires were calculated from line profiles of AFM topography as shown in Fig.~S3.}
\label{fig:Resistivity vs height after annealing}
\end{figure}

Figure~\ref{fig:Resistivity vs height after annealing} shows the dependence of the resistivity of the annealed nanowires on their nominal height.
Each circular marker shows the median value of the resistivity.
The error bars represent the minimum and maximum values.
The lowest measured resistivity is 3.0\,$\mu\Omega$m, observed for a nanowire with a nominal height of 30\,nm.
This value is about one order of magnitude greater than the bulk resistivity of platinum (0.107\,$\mu\Omega$m).
The mean resistivity appears to increase with increasing nominal height.
This is due to the fact that the extracted cross-sectional area is overestimated by the effect of tip convolution (dilation) \cite{villarrubiaAlgorithmsScannedProbe1997}.
This means that the extracted resistivity for shorter nanowires (nominal height $\leq 50$\,nm) is a more accurate estimate of the intrinsic resistivity of the deposited nanowires.

\begin{figure}
\centering
\includegraphics[width=0.99\textwidth]{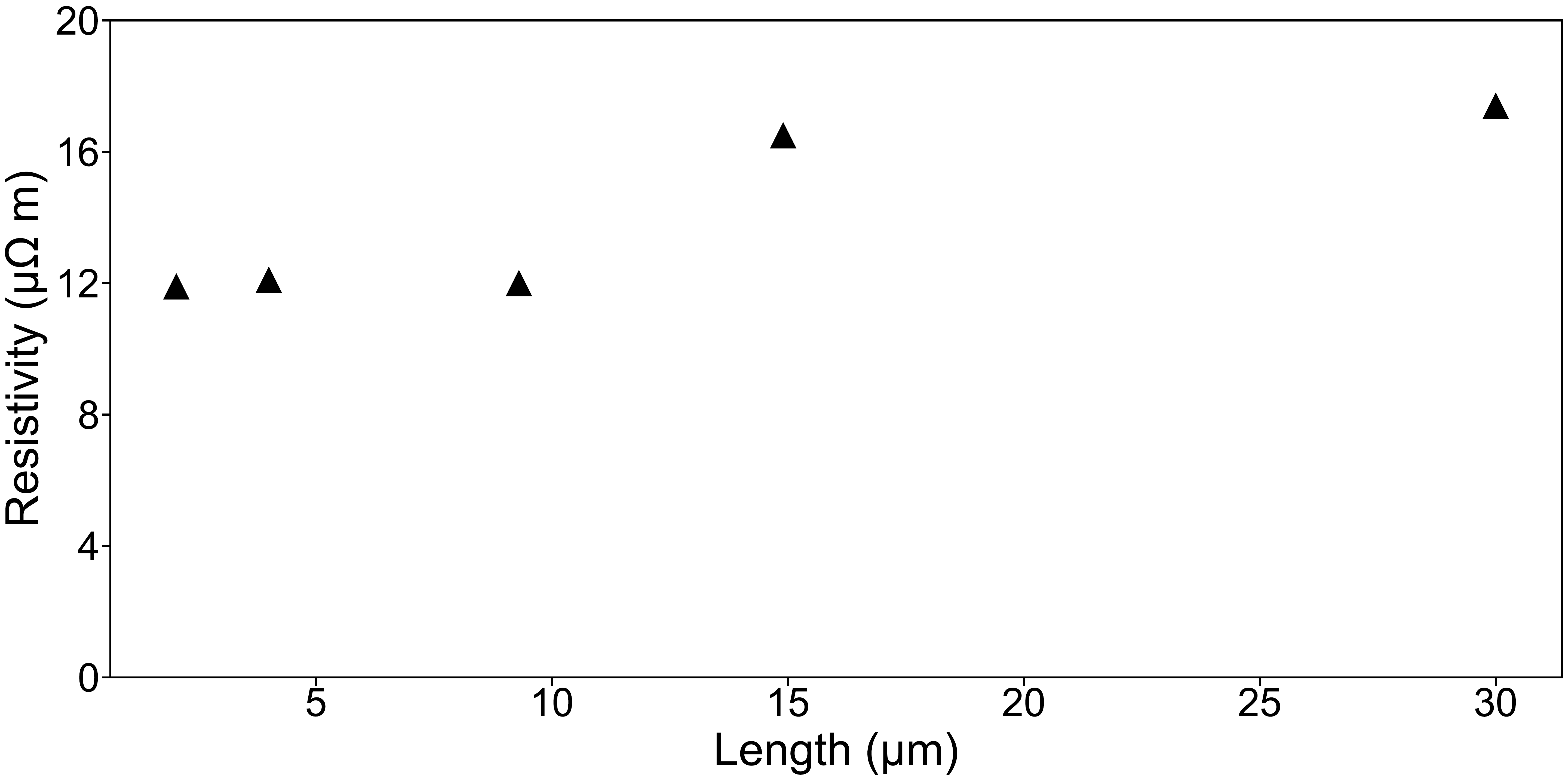}
\caption{Resistivity of annealed nanowires deposited at 5\,kV and 690\,pA with various lengths.
For nanowires with length 5\,$\mu$m or shorter, the pitch value was set to 1\,nm, and for those with length 10\,$\mu$m or longer, it was set to 22.5\,nm.
The measured height of the nanowires was 44\,nm.}
\label{fig:Resistivity VS Length after annealing.png}
\end{figure}

Figure~\ref{fig:Resistivity VS Length after annealing.png} shows the resistivity of the annealed nanowires deposited at 5\,kV and 690\,pA with various lengths.
The measured height of these nanowires was 40\,nm after annealing (nominal height: 133\,nm).
The AFM-measured cross-sectional area was $1.57\times10^{4}~\text{nm}^2$.
This one value was used for all the nanowires with different lengths.
The resistivity is roughly constant over varying nanowire length.
The estimated resistivity is on the order of ten micro-ohm meter, which is two orders of magnitude greater than the bulk resistivity.
The observed deviation from the bulk resistivity is a commonly observed feature in other FEBID-deposited Pt nanowires, as summarized in Table~\ref{table : comparison}, and can be attributed to the granularity of the metallic structures, surface scattering, and the presence of impurities or defects in the nanowires \cite{deteresaOriginDifferenceResistivity2009}.
The lowest measured volume resistivity of 3.0\,$\mu\Omega$m is, to the best of our knowledge, the lowest reported value for FEBID-Pt nanowires.

\subsection{Temperature dependence of resistance of nanowires}
\begin{figure}
\centering
\includegraphics[width=0.99\textwidth]{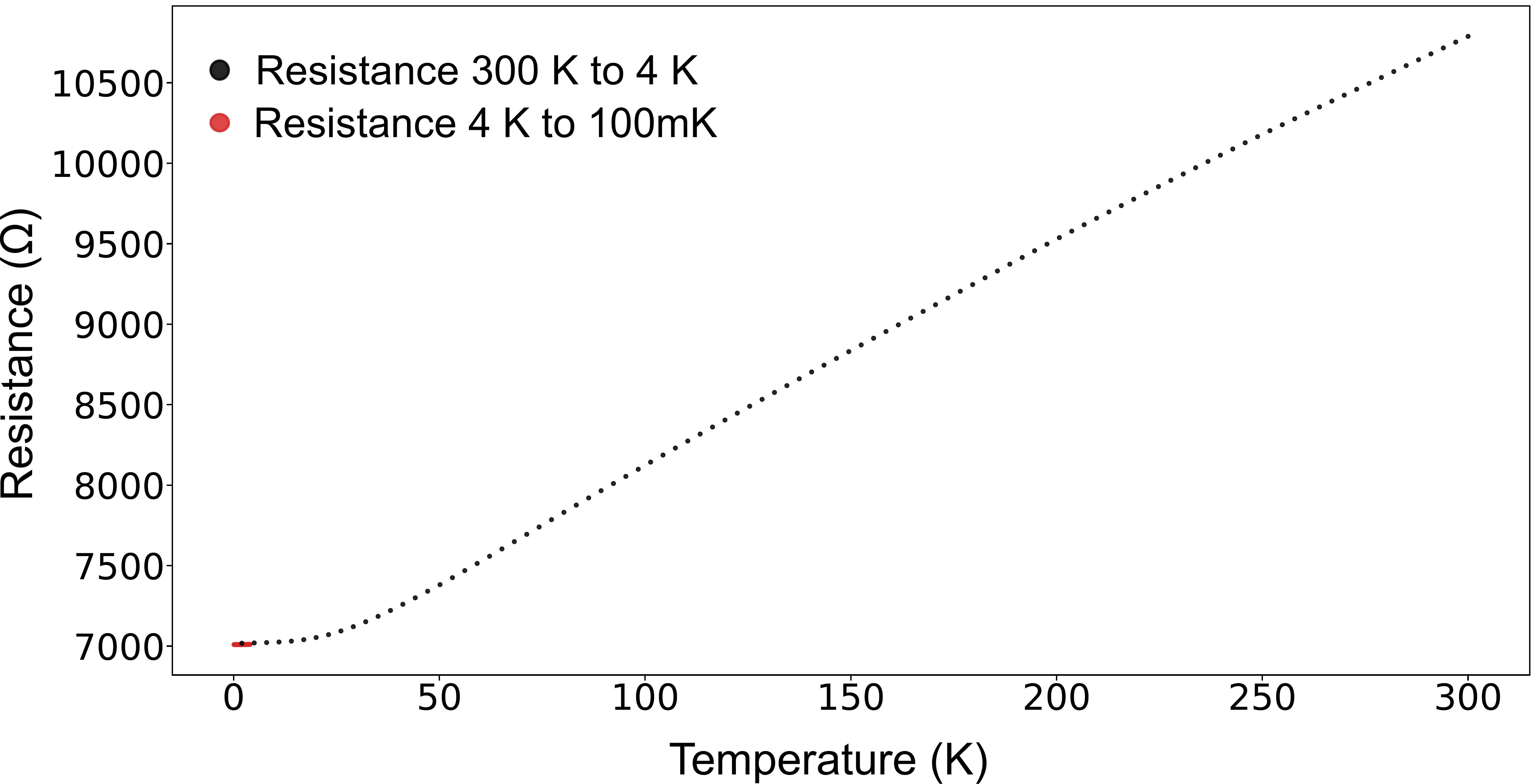}
\caption{Temperature dependence of the resistance of an annealed Pt nanowire deposited at 5\,kV and 690\,pA (nominal height: 133\,nm, measured height: 44\,nm, length: 16.7\,$\mu$m)
 measured from 300\,K to 100\,mK .
A maximum power limit and a current limit of 1\,$\mu$A were set in order to avoid burning the nanowires.}
\label{fig:resistance_vs_T}
\end{figure}

To evaluate the electrical transport properties at cryogenic temperatures, the resistance of an annealed nanowire was measured from 300\,K down to 100\,mK.
Figure~\ref{fig:resistance_vs_T} shows the temperature dependence of the resistance of the annealed nanowire deposited at 5\,kV and 690\,pA (nominal height: 133\,nm, measured height: 44\,nm, length: 16.7\,$\mu$m, the nanowire shown in Fig.~\ref{fig: afm_130nm_ba_aa}(a)).
The resistance of the nanowire decreases with decreasing temperature down to 100\,mK.
This metallic conduction behavior is in contrast to the negative temperature coefficient of resistance observed in Ref.~\cite{fernandez-pachecoPtNanowiresCreated2011}.
We observed the same results with multiple nanowires with various nominal heights
as small as 10\,nm (see Fig.~S6).
Supporing figure Fig.~S7 shows the details of the resistance vs temperature relation between 100\,mK and 4\,K.
Our results demonstrate that these annealed Pt nanowires remain highly conductive and metallic down to millikelvin temperatures.

\section{Conclusion}
Thermal annealing in air at 225\,$^\circ$C reproducibly transforms FEBID-deposited Pt nanowires into highly conductive, 
metallic Pt nanowires whose transport remains metallic down to 100\,mK.
The treatment reduces the resistance by up to five orders of magnitude while simultaneously decreasing the carbon content by about 90\,\% or more and increasing the Pt:C weight ratio to approximately 85:15, indicating efficient purification without structural fragmentation.
Annealing is accompanied by substantial and systematic shrinkage of the nanowire cross section, with the height reduced to 25\,\% or less of the as-deposited value, the width nearly halved, and the cross-sectional area reduced by about an order of magnitude, yet the nanowires remain continuous.
For nanowires with nominal heights between 2\,nm and 200\,nm and lengths around 2\,$\mu$m, the resistance decreases monotonically with increasing height, and for annealed wires the resistance scales approximately linearly with length, consistent with bulk-like diffusive transport.
The lowest volume resistivity achieved is 3.0\,$\mu \Omega$m, which is, to the best of our knowledge, the lowest reported value for FEBID-Pt nanowires and only about one order of magnitude above bulk Pt, with the remaining enhancement attributable to granularity, surface scattering, and residual impurities.
The observation of metallic conduction in nanowires as thin as about 5–10\,nm from room temperature down to 100\,mK establishes annealed FEBID Pt nanowires as robust building blocks for nanoelectronic circuits operating at millikelvin temperatures.
The combination of direct-write patterning, strong and predictable geometric shrinkage, and ultra-low resistivity after annealing makes these structures particularly promising as gate-defined or proximity leads and interconnects in single-electron transistors and other quantum nanoelectronic devices, and provides a practical route toward further miniaturization of cryogenic nanoelectronics based on FEBID.

\newpage
\ack
We gratefully acknowledge funding from NSF-PREM (DMR-2122041), NSF-CAREER (DMR-2044920), and NSF-MRI (DMR-2117438).
This work is also financially supported by Texas State University.
We acknowledge technical assistance from Dr.~Casey Smith and Dr.~Joyce Anderson of Core Research Operations at Texas State University.

\section*{Data availability}
All data that support the findings of this study are available 
in the Texas State University Dataverse Repository, https://doi.org/10.18738/T8/JFXOWC.
% The datasets generated during and/or analyzed during the current study are available in the
% Texas State University Dataverse Repository.

\section*{Conflict of interest}
The authors have no conflict to disclose.

\section*{References}
\bibliography{FEBID.bib}

@article{gopalRapidPrototypingSiteSpecific2004,
  title = {Rapid {{Prototyping}} of {{Site-Specific Nanocontacts}} by {{Electron}} and {{Ion Beam Assisted Direct-Write Nanolithography}}},
  author = {Gopal, Vidyut and Radmilovic, Velimir R. and Daraio, Chiara and Jin, Sungho and Yang, Peidong and Stach, Eric A.},
  year = 2004,
  month = nov,
  journal = {Nano Lett.},
  volume = {4},
  number = {11},
  pages = {2059--2063},
  publisher = {American Chemical Society},
  issn = {1530-6984},
  doi = {10.1021/nl0492133},
  urldate = {2025-12-19}
}

@article{villarrubiaAlgorithmsScannedProbe1997,
  title = {Algorithms for Scanned Probe Microscope Image Simulation, Surface Reconstruction, and Tip Estimation},
  author = {Villarrubia, J.S.},
  year = 1997,
  month = jul,
  journal = {J. Res. Natl. Inst. Stand. Technol.},
  volume = {102},
  number = {4},
  pages = {425},
  issn = {1044677X},
  doi = {10.6028/jres.102.030},
  urldate = {2025-10-06},
  langid = {english}
}

@article{botmanPurificationPlatinumGold2006,
  title = {Purification of Platinum and Gold Structures after Electron-Beam-Induced Deposition},
  author = {Botman, A. and Mulders, J. J. L. and Weemaes, R. and Mentink, S.},
  year = {2006},
  month = jul,
  journal = {Nanotechnology},
  volume = {17},
  number = {15},
  pages = {3779},
  issn = {0957-4484},
  doi = {10.1088/0957-4484/17/15/028},
  urldate = {2024-10-17},
  langid = {english}
}

@article{deteresaOriginDifferenceResistivity2009,
  title = {Origin of the {{Difference}} in the {{Resistivity}} of {{As-Grown Focused-Ion-}} and {{Focused-Electron-Beam-Induced Pt Nanodeposits}}},
  author = {De Teresa, J. M. and C{\'o}rdoba, R. and {Fern{\'a}ndez-Pacheco}, A. and Montero, O. and Strichovanec, P. and Ibarra, M. R.},
  year = {2009},
  journal = {J. Nanomater.},
  volume = {2009},
  number = {1},
  pages = {936863},
  issn = {1687-4129},
  doi = {10.1155/2009/936863},
  urldate = {2024-06-20},
  copyright = {Copyright {\copyright} 2009 J. M. De Teresa et al.},
  langid = {english}
}

@article{durraniElectronTransportRoom2017,
  title = {Electron Transport and Room Temperature Single-Electron Charging in 10 Nm Scale {{PtC}} Nanostructures Formed by Electron Beam Induced Deposition},
  author = {Durrani, Z. A. K. and Jones, M. E. and Wang, C. and Scotuzzi, M. and Hagen, C. W.},
  year = {2017},
  month = nov,
  journal = {Nanotechnology},
  volume = {28},
  number = {47},
  pages = {474002},
  publisher = {IOP Publishing},
  issn = {0957-4484},
  doi = {10.1088/1361-6528/aa9356},
  urldate = {2023-11-01},
  langid = {english}
}

@article{fanAnnealingEffectPlatinumincorporated2014,
  title = {Annealing Effect of Platinum-Incorporated Nanowires Created by Focused Ion/Electron-Beam-Induced Deposition},
  author = {Fang, Jing-Yue and Qin, Shi-Qiao and Zhang, Xue-Ao and Liu, Dong-Qing and Chang, Sheng-Li},
  year = {2014},
  month = jun,
  journal = {Chinese Phys. B},
  volume = {23},
  number = {8},
  pages = {088111},
  issn = {1674-1056},
  doi = {10.1088/1674-1056/23/8/088111},
  urldate = {2024-01-15},
  langid = {english}
}

@incollection{fernandez-pachecoPtNanowiresCreated2011,
  title = {Pt--{{C Nanowires Created}} by {{FIBID}} and {{FEBID}}},
  booktitle = {Studies of {{Nanoconstrictions}}, {{Nanowires}} and {{Fe}}{$_{3}$}{{O}}{$_4$} {{Thin Films}}: {{Electrical Conduction}} and {{Magnetic Properties}}. {{Fabrication}} by {{Focused Electron}}/{{Ion Beam}}},
  author = {{Fernandez-Pacheco}, Amalio},
  editor = {{Fernandez-Pacheco}, Amalio},
  year = {2011},
  pages = {99--127},
  publisher = {Springer},
  address = {Berlin, Heidelberg},
  doi = {10.1007/978-3-642-15801-8_5},
  urldate = {2024-10-12},
  isbn = {978-3-642-15801-8},
  langid = {english}
}

@article{huthFocusedElectronBeam2012,
  title = {Focused Electron Beam Induced Deposition: {{A}} Perspective},
  shorttitle = {Focused Electron Beam Induced Deposition},
  author = {Huth, Michael and Porrati, Fabrizio and Schwalb, Christian and Winhold, Marcel and Sachser, Roland and Dukic, Maja and Adams, Jonathan and Fantner, Georg},
  year = {2012},
  month = aug,
  journal = {Beilstein J. Nanotechnol.},
  volume = {3},
  number = {1},
  pages = {597--619},
  publisher = {Beilstein-Institut},
  issn = {2190-4286},
  doi = {10.3762/bjnano.3.70},
  urldate = {2024-10-12},
  copyright = {{\copyright} 2012 Huth et al; licensee Beilstein-Institut.},
  langid = {english}
}

@article{huthFocusedElectronBeam2018c,
  title = {Focused Electron Beam Induced Deposition Meets Materials Science},
  author = {Huth, M. and Porrati, F. and Dobrovolskiy, O. V.},
  year = {2018},
  month = jan,
  journal = {Microelectronic Engineering},
  volume = {185--186},
  pages = {9--28},
  issn = {0167-9317},
  doi = {10.1016/j.mee.2017.10.012},
  urldate = {2024-10-12}
}

@article{muldersElectronBeamInduced2010,
  title = {Electron Beam Induced Deposition at Elevated Temperatures: Compositional Changes and Purity Improvement},
  shorttitle = {Electron Beam Induced Deposition at Elevated Temperatures},
  author = {Mulders, J. J. L. and Belova, L. M. and Riazanova, A.},
  year = {2010},
  month = dec,
  journal = {Nanotechnology},
  volume = {22},
  number = {5},
  pages = {055302},
  issn = {0957-4484},
  doi = {10.1088/0957-4484/22/5/055302},
  urldate = {2024-10-12},
  langid = {english}
}

@article{porratiTuningElectricalConductivity2011a,
  title = {Tuning the Electrical Conductivity of {{Pt-containing}} Granular Metals by Postgrowth Electron Irradiation},
  author = {Porrati, F. and Sachser, R. and Schwalb, C. H. and Frangakis, A. S. and Huth, M.},
  year = {2011},
  month = mar,
  journal = {J. Appl. Phys.},
  volume = {109},
  number = {6},
  pages = {063715},
  publisher = {American Institute of Physics},
  issn = {0021-8979},
  doi = {10.1063/1.3559773},
  urldate = {2022-11-08}
}

@article{primaDirectwriteSingleElectron2019,
  ids = {primaDirectwriteSingleElectron2019a,primaDirectwriteSingleElectron2019b},
  title = {Direct-Write Single Electron Transistors by Focused Electron Beam Induced Deposition},
  author = {Prima, Giorgia Di and Sachser, Roland and Trompenaars, Piet and Mulders, Hans and Huth, Michael},
  year = {2019},
  journal = {Nano Futur.},
  volume = {3},
  number = {2},
  pages = {025001},
  publisher = {IOP Publishing},
  issn = {2399-1984},
  doi = {10.1088/2399-1984/ab151c}
}

@article{Sachser2011PRL,
  author  = {Sachser, Roland and Porrati, Fabrizio and Schwalb, Christian H. and Huth, Michael},
  title   = {Universal Conductance Correction in a Tunable Strongly Coupled Nanogranular Metal},
  journal = {Physical Review Letters},
  volume  = {107},
  number  = {20},
  pages   = {206803},
  year    = {2011},
  doi     = {10.1103/PhysRevLett.107.206803}
}
\bibliographystyle{unsrt}

\end{document}